# Time-space encoded readout for noise suppression and scalable scanning in optically active solid-state spin systems

**Joachim P. Leibold**†[1,2,3], **Nick R. von Grafenstein**†[1], **Xiaoxun Chen**[1], **Linda Müller**[1], **Karl D. Briegel**[1,3], **Dominik B. Bucher**[1,3]*

[1] Technical University of Munich, TUM School of Natural Sciences, Department of Chemistry, Munich, Germany
[2] Technical University of Munich, TUM School of Natural Sciences, Department of Physics, Munich, Germany
[3] Munich Center for Quantum Science and Technology (MCQST), Schellingstr. 4, D-80799

†These authors contributed equally to this work
*Corresponding author: dominik.bucher@tum.de

**Abstract**

Optically active solid-state spin systems play an important role in quantum technologies. We introduce a new readout scheme, termed "Time to Space" (T2S) encoding which decouples spin manipulation from optical readout both temporally and spatially. This is achieved by controlling the spin state within a region of interest, followed by rapid scanning of the optical readout position using an acousto-optic modulator. Time tracking allows the optical readout position to be encoded as a function of time. Using nitrogen-vacancy (NV) center ensembles in diamond, we first demonstrate that the T2S scheme enables correlated experiments for efficient common mode noise cancellation in various nano- and microscale sensing scenarios. In the second example, we show highly scalable multi-pixel imaging that does not require a camera and has the potential to accelerate data acquisition by several hundred times compared to conventional scanning methods. We anticipate widespread adoption of this technique, as it requires no additional components beyond those commonly used in optically addressable spin systems.

**Introduction**

Optically active solid-state spins are promising candidates for quantum information and sensing applications[1–3]. Color centers in diamond[4], silicon carbide[5], or Van der Waals materials such as hexagonal boron nitride (hBN)[6] are prime examples of such systems. The spin state of these atom-like systems can be prepared, manipulated, and read out by a combination of microwave (MW) and optical excitation pulses (Fig. 1(a)). In such experiments, a measurement spot is defined by a focused laser spot that illuminates the spin (ensemble) (Fig. 1(b)). This leads to a polarization of the spin (ensemble) due to spin-dependent optical transitions. Afterward, the evolution of the quantum state is coupled to a physical quantity, such as the magnetic field, that influences the evolution of the spin state using microwave pulse sequences[7]. These sequences are constructed to project the evolution outcome to the spin state of the system, which can be read out optically with a contrast decaying with the $T_1$ relaxation time of the spin system. In order to probe several spins (ensembles) spatially resolved, either scanning approaches[8,9] or widefield imaging[10–13] has been used so far.

Here, we propose and demonstrate a new scheme called Time-to-Space (T2S) encoding that decouples microwave (MW) manipulation from optical initialization and readout. We apply uniform spin control across the area of interest, followed by rapid, time-tracked local optical excitation, which links the readout time to specific positions (Fig. 1(c)). Importantly, the scanning time to measure the different readout positions must be faster than the $T_1$ spin relaxation time of the system. This approach allows for multiple spatially resolved readouts from a single MW pulse sequence. We show the straightforward implementation of the T2S scheme in a typical experimental setup used for optically active spin defects, and demonstrate two powerful applications: (i) common-mode noise rejection for increased signal-to-noise ratio (SNR) in various scenarios, and (ii) early applications in multi-pixel imaging sequences with the potential to massively accelerate data acquisition in the future.

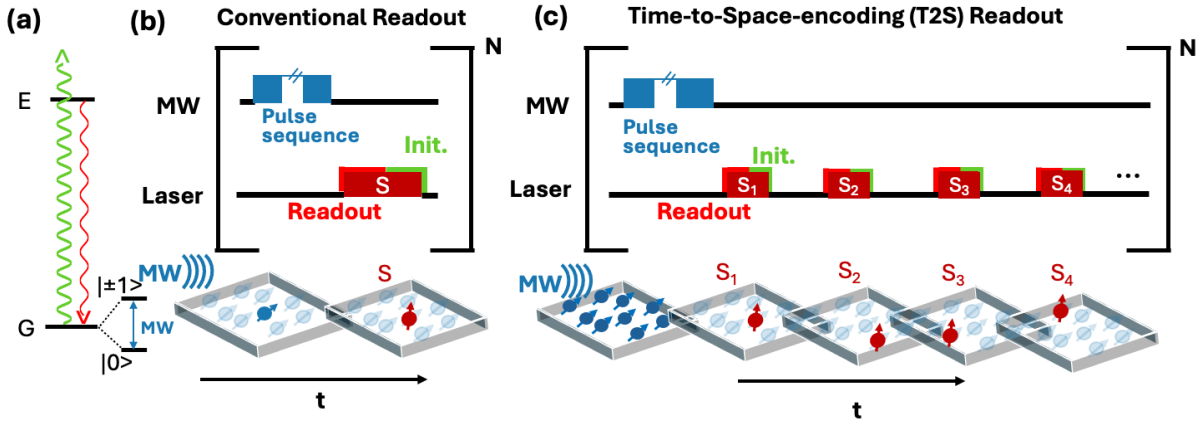

**Figure 1. Comparison of traditional optical readout and the Time-to-Space (T2S) encoding readout for solid-state spin systems.** (a) Energy level scheme for optically active spin defects. The spin states are manipulated using a microwave (MW) pulse sequence (blue), while the initialization (green) and readout (red) of the spin state are performed optically (see also (b)). The microwave protocols are designed to encode a physical observable (that couples to the system) and store it in the spin state, which is subsequently optically read out. (b) Conventional readout: The area of interest, defined by the initialization and readout area (controlled by the laser spot position S) is manipulated using microwave sequences. For averaging purposes, the sequence is repeated for a certain number of times (N). (c) In the T2S scheme, the MW pulse sequence manipulates spins over a region of interest while the optical readout is scanned to achieve multiple ($S_1$ to $S_4$) local readouts during the lifetime of the spin system (T1). The pulse spacing and duration timings do not correspond to the final implementation. This eliminates noise correlations between different readouts and speeds up data acquisition in the typical case where the MW pulse sequence is longer than the optical pulses.

## Results

**Experimental implementation.** The time-to-space encoding scheme (T2S) can easily be implemented in experiments that are already capable of performing pulsed optical detected magnetic resonance (ODMR) measurements on solid-state spin systems (Fig. 2, Material and Methods). For the experiments presented here, we use ensembles of nitrogen-vacancy (NV) centers. First, it is essential to achieve homogeneous microwave (MW) control over the target readout area on the diamond surface. The microwaves couple to the spin states and allow for coherent manipulation of their quantum state[7]. Since the evolution of the quantum state depends on the microwave pulse amplitude, it is important to keep it homogeneous over an area that is supposed to be read out. A simple loop antenna, as used for the experiments presented in this Letter, is sufficient for a few spot measurements[14]. To increase the number of points/sampled area, a resonator can be used to obtain a homogeneous control field over a larger area[15]. The encoded information begins to decay at the end of the pulse sequence on the order of a few milliseconds for the NV center ensemble at room temperature[16]. The spin state can be determined by interrogating the fluorescence upon green laser excitation, which is focused on the NV ensemble and read out on a (large area) photodetector (PD). Second, we need fast switchable read-out positions (spots) with scanning speeds much faster than the $T_1$ relaxation time. Classical scanning approaches such as piezo, galvanic mirrors, or spatial light modulators have scan times of ~ 100 μs, which is too slow compared to typical $T_1$ times[17,18]. Here, we exploit the capabilities of acousto-optical modulators (AOMs) that are designed to function as fast, inertia-free beam steering devices and are commonly used in ODMR experiments. AOMs rapidly deflect the incoming laser beam by an angle θ when driven by radio frequencies (RF). Typically, a single RF tone is applied, and the resulting first-order diffraction of the laser beam is coupled into the optical path of the setup to obtain fast laser pulses for pulsed ODMR experiments[14,19]. Typical rise times are on the nanosecond timescale. The angle of the first-order diffraction θ is, however, not fixed but can be varied by the RF drive frequency [20,21]. A different laser angle results in different laser spots on the diamond surface (see supporting information). These scanning capabilities can be easily implemented by replacing the typically used single-tone frequency drive with a train of pulses of different frequencies (two shown in Fig. 2, with RF frequencies $f_1$ and $f_2$, see Materials and Methods). With this modification, we are able to address multiple (read-out) positions, which include reading out the fluorescence intensity for a few microseconds, moving (~ 100 ns) the laser spot to the next position, and repeating this procedure for the following positions. Since each cycle is much shorter than the $T_1$ time of the spin system, multiple spots can be read out. The measurement results (i.e., fluorescence intensity) and their locations are encoded by the applied RF frequency (corresponding position). The acquisition time and the corresponding radiofrequency of the AOM are controlled by a common timer.

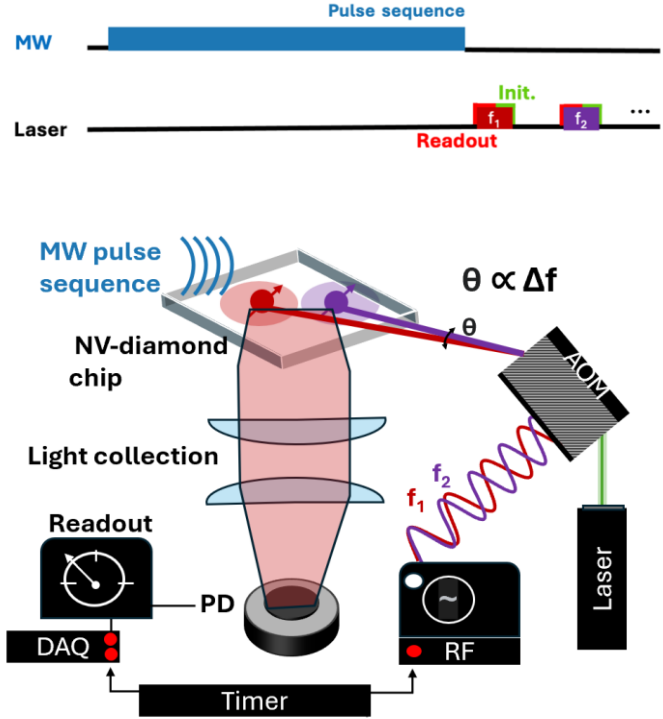

**Figure 2. Experimental implementation of the time-to-space (T2S) encoded readout.** The experimental setup and pulse sequence are similar to a typical pulsed optically detected magnetic resonance (ODMR) setup. Homogeneous microwave fields (blue) are used to deliver pulse sequences for common spin control over a region of interest. For the T2S scheme, we replace the single-tone RF input of the acousto-optic modulator (AOM) with multiple RF frequencies (e.g., in this example two frequencies $f_1$ and $f_2$) that control the deflection angle and allow for addressing different laser spot positions on the NV-diamond chip. Due to the fast response of the AOM, multiple spots on the diamond can be scanned during the relaxation time of the spin system. Their respective states are read out through the spin-state-dependent fluorescence on a photodetector (PD). The voltage of the PD is recorded using a data acquisition (DAQ) unit, and the link between the read-out and the position is achieved by a trigger (Timer) that controls the RF frequency and the acquisition time.

**T2S scheme for correlated measurements and noise reduction.** An important application of the T2S scheme is a correlated two-spot referenced measurement configuration where one of the spots is used as an internal reference (Fig. 3(a)). Applying the same measurement protocol (i.e., pulse sequence) to two points in space (or time) allows for subtraction of the contributions that act similarly on both defined spots, such as background signals or common mode noise induced by the MW pulse sequences (Fig. 3(b)) [22–25]. This is particularly powerful when a sample can be localized at one of the positions, while the other serves as a reference. In our first example, we demonstrate this application to $T_1$ relaxation measurements of shallow NVs, which are strongly dependent on diamond surface properties[26–29]. We cover half of the diamond with paramagnetic $Mn^{2+}$ ions, which strongly reduces the $T_1$ relaxation of the NVs compared to a clean spot (reference) (Fig. 3(c))[30,31]. Common influences such as surface effects, charge state artifacts[32], and drifts, e.g., of the magnetic field, can be tracked and removed with this approach. In addition to the direct correlation of the readouts, scalable scanning of different spots can be achieved since the $T_1$ pulse sequence is much longer than the optical readout.

In our second example, we show how strong background signals (e.g. from the (nuclear) spin bath) in nanoscale NV-NMR experiments [9,33–37] can be mitigated. In NV-NMR experiments, dynamical decoupling sequences (such as the XY8-N) are used to probe different frequency windows by sweeping the π pulse spacing (t) [14,38] (Fig. 3(b)). Matching the spacing with a certain (Larmor) frequency of a target spin results in a stronger coherence decay over the background (peak dip). In experiments with ensembles of NVs on a clean diamond, we observe several peaks that we associate with the diamond spin bath[39], including strong couplings and spurious harmonics[40]. These unwanted signals obscure the spectrum and make the detection of target spins at the diamond surface difficult or impossible. To demonstrate the capabilities of the novel T2S scheme, we implement it for an XY8-4 pulsed dynamical decoupling sequence with two spots on a clean diamond sample (Fig.3(b)). First, we demonstrate the background rejection capabilities for nanoscale NV ensembles using a diamond chip with shallowly implanted NV centers (~ 5nm deep, details see Materials and Methods). The resulting signal of the XY8-4 sequence for the two spots is shown in Fig. 3(d), as well as the curve obtained by subtracting the two data sets. One can see how strongly the curves for the two spots are correlated and how our approach can remove this background. We obtain a threefold decrease in noise after subtracting the two measured data sets (Spot 1 and Spot 2, see supporting information). Thus, we observe how the influence of the NV ensemble on the diamond environment and surface can be compensated. In the case of detecting a sample on the diamond surface, we observed a variation in contrast and $T_2$ time, which made direct noise cancellation difficult. Adding a surface termination layer may solve this problem in the future[9,41,42].

In our final example, we demonstrate the cancellation of noise induced by the pulse sequence. This is a common problem in high-sensitivity experiments with large NV ensembles, where the sensitivity is not limited by photon shot noise but may be constrained by other technical factors[43,44] such as power fluctuations[15] or MW phase noise[23]. In our example, we use a highly sensitive NV ensemble (e.g., ~ 10 μm thick NV layer, in combination with the coherently averaged synchronized readout (CASR) sequence for NMR signal detection, which has been described in detail previously[45–47]. The use of a microfluidic structure[48] and the calibrated positioning of the T2S readout spots allows us to illuminate one spot beneath the sample (water) and the second one beneath the glass structure for reference (Figure 3(e)). Figure 3(f) shows the NMR spectra of both spots, where we observe the 1H NMR water peak at ~ 5000 Hz and a calibration signal at ~ 1000 Hz. Importantly, the noise in both readouts is strongly correlated (Spot 1 and 2, see inset Fig. 3(f)). Subtraction of the data sets reduces the noise by 60%, which we attribute to, among other things, the cancellation of MW phase noise[23] (see materials and methods). We note that the T2S scheme allows the use of affordable signal sources in the future, where even higher improvements are expected. Interestingly, not only does the noise floor decrease, but also the intensity of the NMR peak increases (Fig. 3(g)). This is caused by the direct subtraction of the time domain data and the fact that the phases of the signal above and beside the sample are opposite [11,49] (Supporting Information). Subtracting this phase-shifted real-time data leads to a higher contrast in the time domain data and, hence, a higher signal amplitude in the Fourier-transformed frequency domain. Together with the reduction of the noise floor, this leads to an improvement of the signal-to-noise ratio of about three (as in the nanoscale case), which corresponds to almost an order of magnitude reduction of the sampling time.

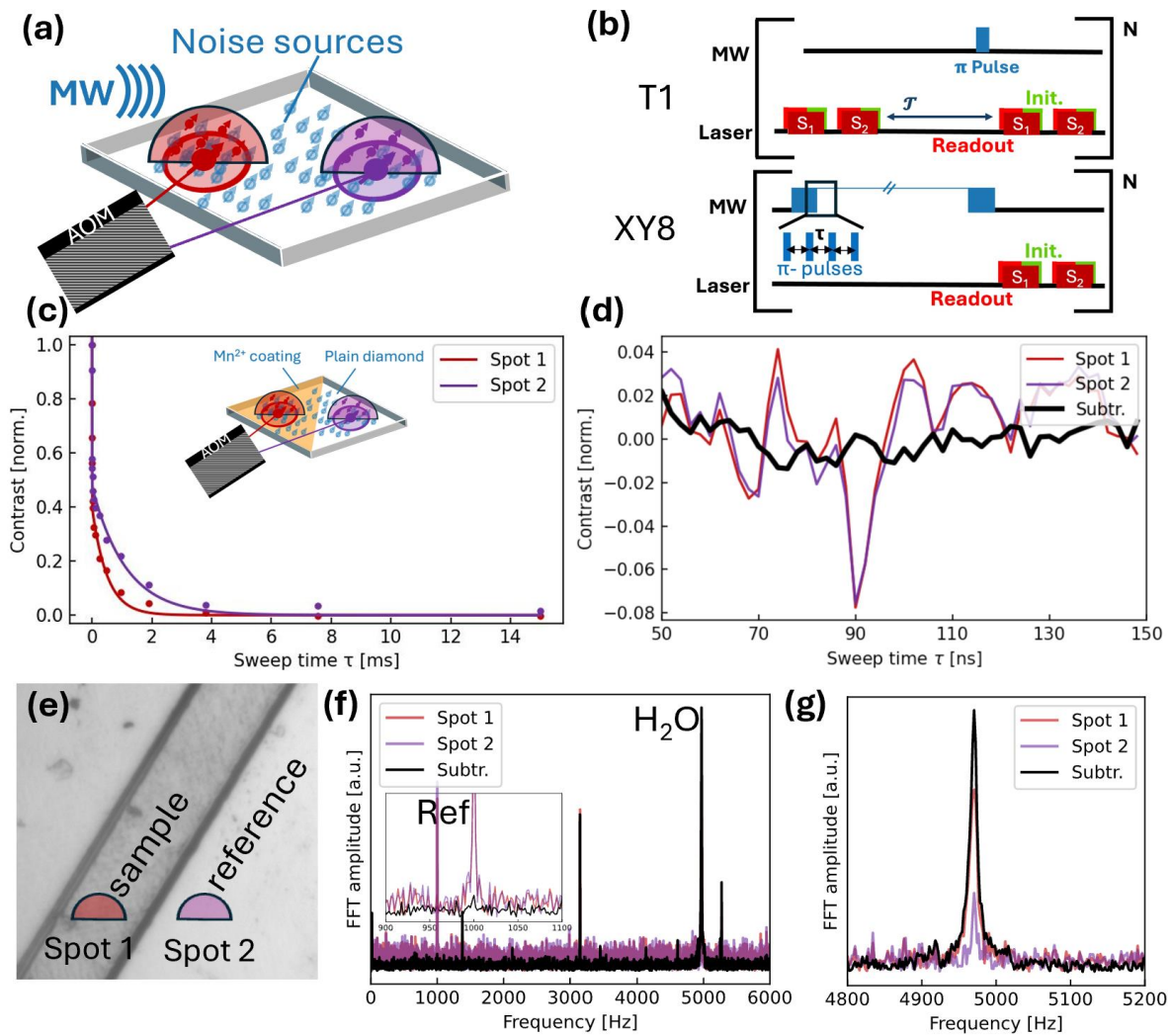

**Figure 3. T2S scheme for correlated measurements and noise reduction.** (a) NV-diamond chip with two T2S encoded spots and their respective sensing volumes on the diamond surface. Noise sources acting globally on both T2S readouts can originate from the microwave pulse sequence or the spin bath of the NV ensemble. (b) Implementation of $T_1$ relaxation and XY8-N dynamical decoupling pulse sequences in the T2S scheme. (c) $T_1$ relaxation experiments with two laser spots with a shallow NV-ensemble diamond half-covered with a paramagnetic sample. (d) XY8-4 measurement with the T2S encoding on a shallow NV-ensemble and subsequent signal subtraction of the obtained signals (Subtr., black). (e) Image of a microfluidic channel separating a sample containing volume from a reference readout (glass). (f) NV-NMR spectrum of the sample and reference position as well as the subtracted data set. The inset shows a zoom into the noise floor where the correlation of the noise can be seen. (g) Zoom on the $^1$H NMR water signal. The (time domain) subtracted data (Ref.) does not only show less noise but also a higher signal, resulting in a 3x increased SNR. This improved contrast is due to the different phases of the NMR signals.

**T2S scheme for scalable scanning.** The two-spot T2S readout can be easily extended to multiple spots, enabling not only noise cancellation but also scalable scanning for imaging applications. This is especially powerful when the pulse sequence is much longer than the optical readout. We have implemented T2S encoding with a four-spot readout sequence for our microscale NV-NMR applications (Figure 4(a)), and Figure 4(b) shows a false-color image of the superimposed spots recorded in the experiment. We use the same CASR sequence and water-filled microfluidic channel as in the previous experiment. The spectrum for each spot ($f_1$-$f_4$) are shown in (Fig. 4(c), top). The $^1$H water signals at ~ 3200 Hz have been normalized to a calibration signal for comparability. Figure 4(c) bottom shows the FFT amplitude (red dots) of the water signal over the channel cross-section. The relative amplitudes of the NMR signals are in good agreement with our simulated data over the entire channel section (blue line). The simulated signal amplitude takes into account the geometric dependence of the signal strength on the NV centers, introduced by the dipolar coupling origin of the signal[11,49]. Further details on the simulation can be found in the Supporting Information.

Importantly, the number of spots can be significantly increased in the future by adapting the optics to meet specific needs (such as scanning area, laser spot size, etc.). The scheme is also not limited to a one-dimensional implementation. By using two perpendicular crystals, a two-dimensional array of spots can be achieved which enables fast imaging without cameras[11] (Supporting Information). To estimate the potential acceleration over conventional scanning approaches, we assume that the full T2S encoding readout time is equal to the pulse sequence duration. Typical optical initialization and readout can be achieved within a ~2 μs laser pulse[50]; an additional 0.5 μs is required to move the laser spots between two positions. Significant increases in the number of readouts are anticipated for long pulse sequence durations. For instance, in a pulse sequence constrained by $T_2$ (100 μs), up to 40 readouts can be achieved, while a sequence limited by $T_1$ (1 ms) can yield up to 400 readouts per sequence, with only a twofold increase in the acquisition time (Fig. 4d).

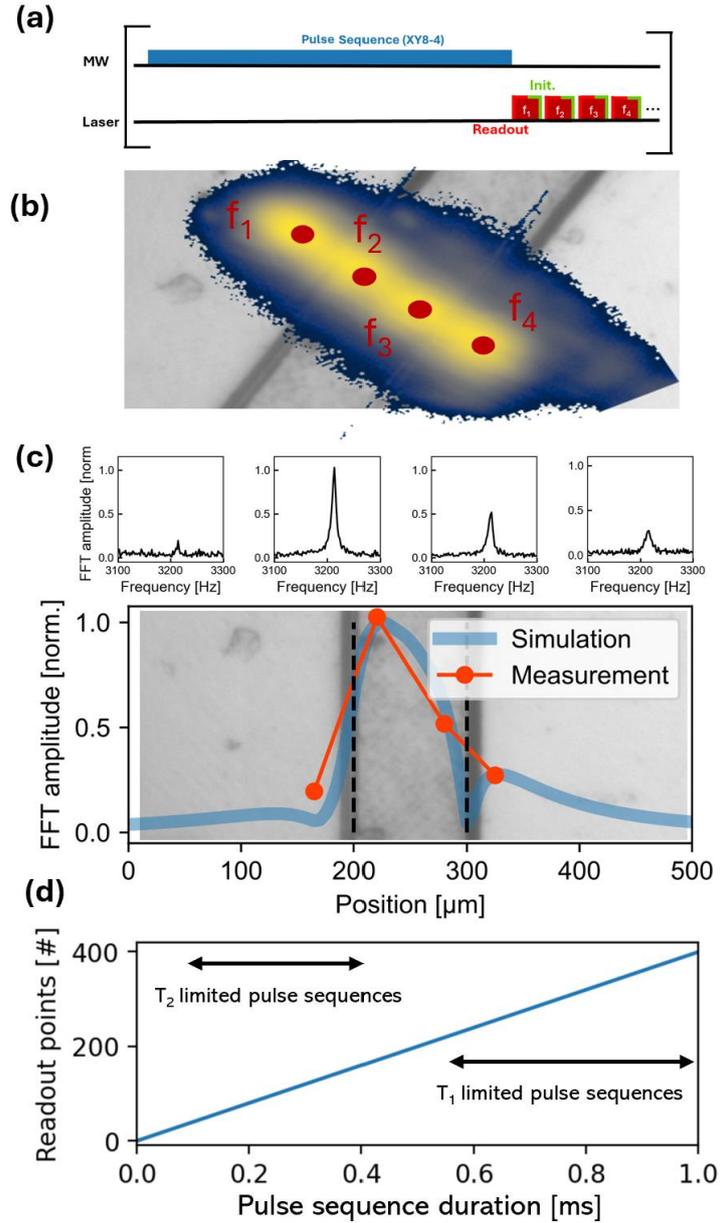

**Figure 4. Scalable scanning application for imaging.** (a) T2S scheme for four readout spots. (b) Overlayed image of the channel and the four spatially resolved laser spots (false color). (c) $^1$H NMR spectra of the four individual readouts defined by the respective AOM driving frequency tone ($f_1$ to $f_4$). All spectra have been normalized to a calibration reference signal. $^1$H NMR signal amplitudes (red dots) and the simulated NMR amplitude (blue) plot as a function of the channel position (bottom). (d) Scalability of the T2S scheme. Number of readouts per pulse sequence as a function of its duration assuming that i) the full T2S readout time is equal to the pulse sequence duration and ii) the duration of 2.5 μs per readout (incl. 2 μs laser pulse and 0.5 μs scanning).

**Conclusion and Outlook**

We have introduced a novel readout scheme for optically active solid-state spin systems (T2S) that decouples spin control from optical initialization and readout, both in time and space. We have demonstrated that this approach can be used i) to cancel efficiently common mode noise sources in nanoscale and microscale sensing applications and ii) to enable a fast few-pixel scalable scanning method, which can lead to massive acceleration of the data acquisition. Importantly, the scheme relies on the fast scanning capabilities of AOMs, which are state of the art in all solid-state spin defects-based experiments and allow for easy implementation. Our work focused on NV-diamond ensembles quantum sensing applications, but we expect widespread applications also for single spin defect experiments, where multiple (~ several 100's) NVs can be probed simultaneously[51–54]. Also, this scheme can be directly applied to nanodiamond relaxation experiments[55,56] or other emerging spin defects, such as in hBN[57,58] or silicon carbide. Due to the wide applicability, straightforward implementation, and drastic improvements in terms of scanning scalability, we expect a broad application of the T2S scheme in the future.

**Note:** During the submission process of our manuscript, we became aware of References [59,60], which employ a related scanning method for reading out multiple single NV centers on a camera.

**Acknowledgements**

**Funding.** This project has been funded by the Bayerisches Staatministerium für Wissenschaft und Kunst through project IQSense via the Munich Quantum Valley (MQV), by the Federal Ministry of Education and Research (BMBF) as part of the VIP+ validation funding program (03VP10350) and the European Research Council (ERC) under the European Union's Horizon 2020 research and innovation programme (Grant Agreement No. 948049). The authors acknowledge further support by the DFG under Germany's Excellence Strategy–EXC 2089/1-390776260 and the EXC-2111 390814868.

**Author Contributions.** D.B.B. conceived and supervised the study. J.P.L., N. v.G, and L.M. performed the experiments and analyzed the data. X. C. performed optics simulations and guided the optics design. K.D.B. performed the NMR simulations. J.P.L. and D.B.B. wrote the manuscript with inputs from all authors.

**Competing Interests.** D.B.B. and J.P.L. have filed a patent application for the T2S readout. All other authors have no competing interests.